\documentclass{aa}
\usepackage{graphicx}
\begin{document}
\title{Type Ia supernova SN 2003du: optical observations}
\author{G.C. Anupama \inst{1}, D.K. Sahu \inst{1,2} \and J. Jose\inst{1,2}}
\offprints{G.C. Anupama}
\institute{Indian Institute of Astrophysics, II Block Koramangala, Bangalore 
560 034, India\\
\email{gca@iiap.res.in}
\and
Center for Research and Education in Science \& Technology, Hosakote\\
\email{dks@crest.ernet.in}}

\date{Received: 19/07/2004; accepted: 15/09/2004}

\abstract{$UBVRI$ photometry and optical spectra of type Ia supernova SN 2003du
obtained at the Indian Astronomical Observatory for nearly a year since
discovery are presented. 

The apparent magnitude at maximum was $B=13.53\pm 0.02$~mag, and the 
colour $(B-V)=-0.08\pm 0.03$~mag. The luminosity decline rate, 
$\Delta m_{15}(B)=1.04\pm 0.04$~mag indicates an absolute $B$ magnitude at 
maximum of $M_B^{\mathrm{max}}=-19.34\pm 0.3$~mag and the distance modulus to 
the parent galaxy as $\mu=32.89\pm 0.4$.
The light curve shapes are similar, though not identical, to those of SNe 1998bu
and 1990N, both of which had luminosity decline rates similar to that of
SN 2003du and occurred in spiral galaxies.
The peak bolometric luminosity indicates that $\sim 0.9\,{\rm{M}}_\odot$ mass
of $^{56}$Ni was ejected by the supernova.
The spectral evolution and the evolution of the Si II and Ca II absorption
velocities closely follows that of SN 1998bu, and in general, is within the
scatter of the velocities observed in normal type Ia supernovae.

The spectroscopic and photometric behaviour of SN 2003du is quite typical for
SNe Ia in spirals.

A high velocity absorption component in the Ca II (H \& K) and IR-triplet 
features, with absorption velocities of $\sim 20\,000$~km s$^{-1}$ and
$\sim 22\,000$~km s$^{-1}$ respectively, is detected in the pre-maximum spectra
of days $-11$ and $-7$.

}

\titlerunning{Type Ia supernova 2003du}
\authorrunning{Anupama et al.}

\maketitle

\section{Introduction}

The type Ia supernovae (SNe Ia) are widely believed to be the result of 
combustion of a degenerate white dwarf (Hoyle \& Fowler \cite{h60}). The most 
likely models are: (1) an explosion of a CO white dwarf with a mass close to 
the Chandrasekhar limit, which accretes matter through Roche lobe overflow 
from an evolved companion star, (Whelan \& Iben \cite{wi73}) or (2) explosion 
of a rotating configuration formed from the merging of two low-mass white 
dwarfs, following the loss of angular momentum due to gravitational radiation 
(Webbink \cite{w84}; Iben \& Tutukov \cite{i84}; Paczy\'nski \cite{p85}; see
Livio \cite{l01} for a review). From 
the observed spectral and light curve properties, the first scenario appears 
to be the most likely candidate for a majority of SNe Ia. In both scenarios, a 
certain amount of material associated with the mass transfer is likely to 
remain in the system at the time of explosion. The interaction of the supernova 
with this material may lead to a shell structure in the ejecta (Khokhlov, 
M\"uller \& H\"oflich \cite{k93}; Gerardy et al.\ \cite{g04}).

The temporal evolution of a supernova's luminosity contains important
information on the physical processes driving the explosion. The observed
bolometric light curves provide a measure of the total output of converted
radiation of SNe Ia, and hence serve as a crucial link to theoretical models
of the explosion and evolution. The peak luminosity is directly linked to the 
amount of radioactive $^{56}$Ni produced in the explosion (e.g.\ Arnett 
\cite{a82}, H\"oflich et al.\ \cite{ho96}, Pinto \& Eastman \cite{pi00}) and 
can be used to test various explosion models.

A majority of SNe Ia belong to a fairly homogeneous class, in both their 
photometric as well as spectroscopic properties. While Branch et al.\ 
(\cite{b93}) estimated 83\% of the SNe Ia in their sample to be ``normal'' 
(SNe Ia that show conspicious absorption features near 6150~\AA\ due to Si II
and near 3750~\AA\ due to Ca II in their spectra near maximum light), a more 
recent study by Li et al.\ (\cite{li01}) indicate 64\% of SNe Ia are ``normal'',
20\% belong to the overluminous type similar to SN 1991T and 16\% belong to the
subluminous type similar to SN 1991bg. The
peak absolute magnitudes of SNe Ia in the $B$, $V$, and $R$ bands correlate
with the decline rate of the immediate post-maximum light curve giving a
photometric sequence from luminous blue events with relatively slow decline
of the light curve to the subluminous red events with rapid decline of the
light curve (e.g.\ Phillips \cite{p93}; Hamuy et al.\ \cite{h96a}; Hamuy et al.
\cite{h96b}; Phillips et al.\ \cite{pet99}). When arranged in the photometric 
sequence, SNe Ia also form a spectroscopic sequence (Branch et al.\ \cite{b93}).
The peak absolute magnitude of SNe Ia is also correlated with the Hubble type
of the parent galaxy. SNe Ia in ellipticals are, on an average, fainter than
SNe Ia in spirals (Della Valle \& Panagia \cite{dp92}; Howell \cite{h01}).
Although a majority of the SNe Ia are normal, some photometric and spectral 
inhomogeneities exist, even amongst these normal SNe Ia. The spectral 
variations are seen to correlate with the
expansion velocity, the effective temperature and the peak luminosity (Nugent
et al.\ \cite{nu95}). The observed diversity indicate that SNe Ia cannot be 
described by a single parameter such as the early light curve decline and that 
the diversity is multidimensional (Hatano et al.\ \cite{ha00}; Benetti et
al.\ \cite{b04}). It is therefore 
important to study individual SNe events covering a wide spectral and temporal 
range. 

We present in this paper the optical photometric and spectroscopic observations
of the type Ia supernova SN 2003du. A description of the observations
and analysis is given in Sec.\ 2, and the light and colour curves presented in
Secs.\ 3 and 4. The absolute magnitude, the bolometric luminosity
and the mass of $^{56}$Ni ejected are discussed in Secs. 5 and 6. Section 7
gives a description of the spectral evolution, which is compared with other
well studied normal SNe Ia.

\section{Observations and Reduction}

The supernova SN 2003du was discovered by Shwartz \& Holvorcem (\cite{s03})
on 2003 April 22.4, located 8.8 arcsec West and 13.5 arcsec South of
the center of the nearby SBdm galaxy UGC 9391. Optical spectra by
Kotak \& Meilke (\cite{k03}) obtained on April 24.06 indicated the supernova to
be of type Ia, about two weeks before maximum. Based on a study of the spectra 
of SN 2003du during the early phases, Gerardy et al.\ (\cite{g04}) report the
detection of a high velocity component in the Ca II infrared triplet near
8000~\AA. This feature exhibits a large expansion velocity, which is nearly
constant between $-7$ and $+2$ days relative to maximum light and
disappears shortly thereafter. Gerardy et al.\ (\cite{g04}) suggest this feature
can be caused by a dense shell formed when circumstellar material of solar
abundance is overrun by the rapidly expanding outermost layers of the SN ejecta.

The observations of SN 2003du with the 2m Himalayan Chandra 
Telescope (HCT) at the Indian Astronomical Observatory (IAO), Hanle, India, 
began on 2003 April 24, two days after the discovery. The HFOSC instrument
equipped with a SITe $2\times 4$~K pixel CCD was used. The central $2\times 2$~K
region used for imaging corresponds to a field of view of 
$10\,{\rm{arcmin}}\times 10\,{\rm{arcmin}}$ at
0.296~arcsec pixel$^{-1}$. More details on the telescope and the instrument 
may be obtained from {\it{http://www.iiap.res.in/$\sim$iao}}.

\subsection{Photometry}
\begin{figure}
\resizebox{\hsize}{!}{\includegraphics{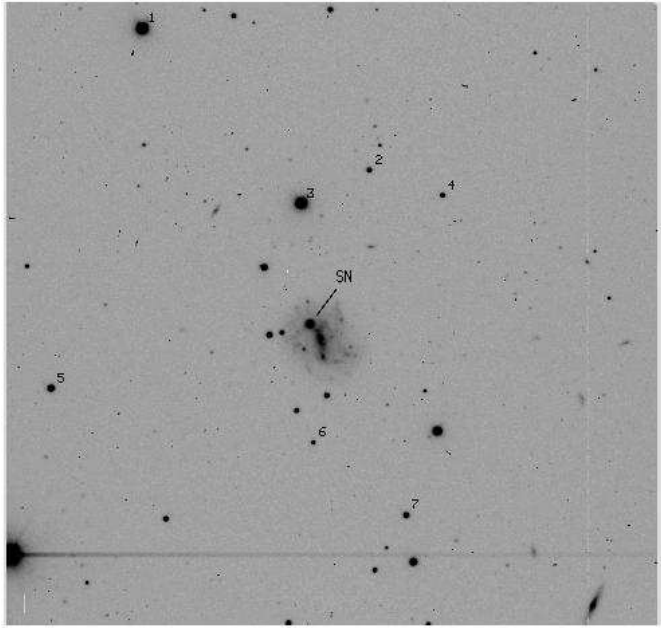}}
\caption{Field of SN 2003du. The field of view is $10^\prime \times 10^\prime$.
The stars used as local standards are listed in Table \ref{tab1}}
\label{fig1}
\end{figure}

$UBVRI$ photometry of SN 2003du were obtained during 24 April 2003 to
21 September 2003 and again in 2004, spanning $-12 - +362$ days since $B$ 
maximum.
The data were bias subtracted and flat field corrected in the standard manner 
using the various tasks under IRAF. Landolt (\cite{la92}) standard regions 
PG0918+029, PG$0942-029$, SA101 (region covering stars 421, 338, 429, 431, 427, 
341) and SA104 (region covering stars 330, 334, 336, 338, 339) were observed on 
24 February 2004 and the regions 
PG0918+029, PG1047+003, PG$1323-086$, PG1530+057 were observed on 1 March 2004 
under photometric conditions. These were used to estimate the atmospheric 
extinction and calibrate the field stars in the supernova region for use as 
secondary calibrators. They yield the following transformation equations:
\begin{eqnarray*}
u & = & U - (0.188\pm 0.02) (U-B) - (3.20\pm 0.02)\\
b & = & B - (-0.050\pm 0.02) (B-V) - (1.01\pm 0.003)\\
v & = & V - (0.042\pm 0.01) (B-V) - (0.634\pm 0.003)\\
r & = & R - (0.064\pm 0.02) (V-R) - (0.721\pm 0.01)\\
i & = & I - (0.017\pm 0.02) (V-I) - (1.01\pm 0.01),\\
\end{eqnarray*}
where $ubvri$ are the instrumental magnitudes corrected for atmospheric 
extinction and $UBVRI$ are the standard
magnitudes. The position of the supernova and the stars in 
the field used as local standards are shown marked in Fig. \ref{fig1}. 
Table \ref{tab1} gives the magnitudes 
of the secondary standards, averaged over the two nights. These magnitudes 
were used to calibrate the data obtained on other nights.

\begin{center}
\begin{table*}
\caption{Magnitudes for the local sequence stars in the field of SN 2003du. The
stars are identified in Fig. \ref{fig1}.}
\begin{tabular}{lrrrrr}
\hline\hline
ID & \multicolumn{1}{c}{U} & \multicolumn{1}{c}{B} & \multicolumn{1}{c}{V} & 
\multicolumn{1}{c}{R} & \multicolumn{1}{c}{I}\\
\hline
1 & $13.80\pm 0.01$ & $13.81\pm 0.02$ & $13.19\pm 0.01$ & $12.80\pm 0.003$ & $12.43\pm 0.003$\\
2 & $18.34\pm 0.01$ & $17.67\pm 0.01$ & $16.30\pm 0.01$ & $15.31\pm 0.02$ & $14.19\pm 0.02$\\
3 & $13.85\pm 0.01$ & $13.92\pm 0.02$ & $13.37\pm 0.01$ & $13.03\pm 0.01$ & $12.72 \pm 0.01$\\
4 & $17.76\pm 0.02$ & $18.04\pm 0.02$ & $17.57\pm 0.01$ & $17.22\pm 0.003$ & $16.90\pm 0.02$\\
5 & $16.53\pm 0.01$ & $16.32\pm 0.03$ & $15.58\pm 0.02$ & $15.15\pm 0.01$ & $14.74\pm 0.01$\\
6 & $18.25\pm 0.04$ & $18.67\pm 0.01$ & $18.20\pm 0.01$ & $17.86\pm 0.01$ & $17.50\pm 0.01$\\
7 & $17.14\pm 0.02$ & $17.05\pm 0.02$ & $16.33\pm 0.02$ & $15.91\pm 0.01$ & $15.51\pm 0.01$\\
\hline
\end{tabular}
\label{tab1}
\end{table*}
\end{center}

Aperture photometry was performed on the local standards using an aperture of
radius 3-4 times that of the FWHM of the seeing profile that was determined 
based on an aperture growth curve. Since the underlying galaxy background at
the location of the supernova is varying, the
supernova magnitudes were obtained using the profile-fitting method, using
a fitting radius corresponding to the FWHM of the seeing profile. The
difference between aperture and profile-fitting magnitudes was obtained using
the standards and this correction was applied to the supernova magnitude.
The supernova magnitudes were calibrated differentially with respect to the
local standards listed in Table \ref{tab1}. The estimated supernova magnitudes 
are listed in Table \ref{tab2}.

\begin{center}
\begin{table*}
\caption{Photometric observations of SN 2003du}
\begin{tabular}{lcrrrrrrr}
\hline\hline
Date & J.D. & \multicolumn{1}{c}{Phase\rlap{*}} & \multicolumn{1}{c}{U} & 
\multicolumn{1}{c}{B} & \multicolumn{1}{c}{V} & \multicolumn{1}{c}{R} & 
\multicolumn{1}{c}{I} & \multicolumn{1}{c}{seeing}\\
     & 2400000+ & \multicolumn{1}{c}{(days)} & & & & & & 
\multicolumn{1}{c}{($^{\prime\prime}$)}\\
\hline
24/4/03 & 52754.2 & $-12.1$ & $15.08\pm 0.03$ & $15.30\pm 0.02$ & 
$15.32\pm 0.02$ & $15.22\pm 0.01$ & $15.30\pm 0.02$ & 1.7\\
25/4/03 & 52755.3 & $-11.0$ & $14.55\pm 0.02$ & $14.82\pm 0.02$ & 
$14.91\pm 0.02$ & $14.75\pm 0.02$ & $14.77\pm 0.03$ & 2.0\\
29/4/03 & 52759.3 & $-7.0$ & $13.44\pm 0.03$ & $13.94\pm 0.02$ & 
$14.06\pm 0.02$ &                   & $14.29\pm 0.02$ & 1.5\\
04/5/03 & 52764.2 & $-2.1$ & $13.07\pm 0.03$ & $13.55\pm 0.02$ & 
$13.64\pm 0.02$ & $13.62\pm 0.01$ & $13.85\pm 0.02$ & 2.1\\
09/5/03 & 52769.3 & $+3.0$ & $13.25\pm 0.02$ & $13.60\pm 0.02$ & 
$13.63\pm 0.01$ & $13.66\pm 0.01$ & $14.04\pm 0.02$ & 1.9\\
11/5/03 & 52771.3 & $+5.0$ & $13.44\pm 0.02$ & $13.72\pm 0.02$ &
                  & $13.73\pm 0.01$ & $14.13\pm 0.01$ & 1.8\\
15/5/03 & 52775.2 & $+8.9$ & $13.64\pm 0.02$ & $13.97\pm 0.02$ & 
$13.84\pm 0.01$ & $13.90\pm 0.01$ & $14.38\pm 0.01$ & 1.7\\
18/5/03 & 52778.3 & $+12.0$ & $14.01\pm 0.02$ & $14.22\pm 0.02$ & 
$14.00\pm 0.02$ & $14.17\pm 0.01$ & $14.57\pm 0.02$ & 1.6\\
19/5/03 & 52779.3 & $+13.0$ & $14.14\pm 0.02$ & $14.35\pm 0.02$ & 
$14.08\pm 0.02$ & $14.23\pm 0.01$ & $14.60\pm 0.01$ & 1.6\\
25/5/03 & 52784.3 & $+18.0$ & $14.76\pm 0.02$ & $14.90\pm 0.02$ & 
$14.36\pm 0.01$ & $14.40\pm 0.01$ & $14.59\pm 0.01$ & 2.0\\
31/5/03 & 52791.4 & $+25.1$ & $15.59\pm 0.02$ & $15.65\pm 0.02$ & 
$14.71\pm 0.01$ & & & 1.9\\
01/6/03 & 52792.3 & $+26.0$ & $15.68\pm 0.03$ & $15.74\pm 0.02$ & 
$14.74\pm 0.02$ & $14.51\pm 0.01$ & $14.42\pm 0.01$ & 1.4\\
23/6/03 & 52814.2 & $+47.9$ &                   &                   & 
$15.72\pm 0.03$ & $15.49\pm 0.01$ & $15.27\pm 0.03$ & 2.4\\
28/6/03 & 52819.3 & $+53.0$ & $16.72\pm 0.02$ & $16.79\pm 0.02$ & 
$15.90\pm 0.02$ & $15.70\pm 0.02$ & & 3.0\\
03/7/03 & 52824.1 & $+57.8$ & $16.85\pm 0.02$ & $16.94\pm 0.02$ & 
$16.05\pm 0.01$ & $15.92\pm 0.02$ & & 2.0\\
01/8/03 & 52853.2 & $+86.9$ & $17.49\pm 0.02$ & $17.34\pm 0.02$ & 
$16.79\pm 0.01$ & $16.75\pm 0.02$ & $16.95\pm 0.02$ & 1.2\\
06/8/03 & 52858.2 & $+91.9$ &                   & $17.46\pm 0.02$ & 
$16.91\pm 0.01$ & $16.90\pm 0.01$ & $17.12\pm 0.01$ & 1.9\\
30/8/03 & 52882.2 & $+115.9$ &                  & $17.87\pm 0.02$ & 
$17.51\pm 0.02$ & $17.61\pm 0.02$ & $17.95\pm 0.02$ & 1.7\\
05/9/03 & 52888.1 & $+121.8$ &                  & $17.95\pm 0.03$ & 
$17.60\pm 0.01$ & $17.75\pm 0.02$ & $18.03\pm 0.02$ & 2.1\\
20/9/03 & 52903.1 & $+136.8$ & $18.71\pm 0.04$ & $18.10\pm 0.03$ & 
$17.90\pm 0.03$ & $18.05\pm 0.03$ & & 1.6\\
21/9/03 & 52904.2 & $+137.9$ &                   & $18.16\pm 0.02$ & 
$17.94\pm 0.02$ & $18.06\pm 0.02$ & $18.14\pm 0.05$ & 1.9\\
24/2/04 & 53060.5 & $+293.7$ &                   & $20.37\pm 0.05$ & 
$20.32\pm 0.04$ & $20.65\pm 0.06$ & $19.90\pm 0.07$ & 1.8\\
01/3/04 & 53066.5 & $+300.2$ &                   & $20.43\pm 0.04$ & 
$20.24\pm 0.05$ & $20.68\pm 0.10$ & $19.88\pm 0.11$ & 2.3\\
14/4/04 & 53110.5 & $+344.2$ &                   &                   &
& $20.70\pm 0.06$ & & 1.9\\
02/5/04 & 53128.3 & $+362.0$ &                   &                   & 
$20.77\pm 0.04$ & $20.66\pm 0.06$ & & 2.0\\
\hline
\multicolumn{9}{l}{\rlap{*}\ \ Relative to the epoch of $B$ maximum JD = 2452766.3}
\end{tabular}
\label{tab2}
\end{table*}
\end{center}

\subsection{Spectroscopy}
Spectroscopic observations of SN 2003du were made on eight occasions beginning
25 April 2003. The journal of observations is given in Table \ref{tab3}. All 
spectra
were obtained at a resolution of $\sim 7$~\AA\ in the wavelength range 
3500-7000~\AA\ and 5200-9200~\AA. The one-dimensional spectra were extracted
from the bias subtracted and flat-field corrected images using the optimal
extraction method. Wavelength calibration was effected using the spectra of
FeAr and FeNe sources. The wavelength calibrated spectra were corrected for
instrumental response using spectra of spectrophotometric standards observed
on the same night and brought to a flux scale. The flux calibrated spectra in
the two different regions were combined, scaled to a weighted mean, to give the 
final spectrum on a relative flux scale.

\begin{table}
\caption{Spectroscopic observations of SN 2003du}
\begin{tabular}{lcrc}
\hline\hline
Date & J.D. & \multicolumn{1}{c}{Phase\rlap{*}} & Range \\
     & 2400000+ & \multicolumn{1}{c}{(days)} & \AA\ \\
\hline
25/04/03 & 52755.5 & $-10.8$ & 3500-7000; 5200-9200\\
29/04/03 & 52759.4 & $-6.9$ & 3500-7000; 5200-9200\\
15/05/03 & 52775.3 & $+9.0$ & 5200-9200\\
19/05/03 & 52779.4 & $+13.1$ & 3500-7000; 5200-9200\\ 
12/06/03 & 52803.2 & $+36.9$ & 3500-7000; 5200-9200\\
01/08/03 & 52853.2 & $+86.9$ & 3500-7000; 5200-9200\\
21/09/03 & 52904.1 & $+137.8$ & 3500-7000; 5200-9200\\
22/09/03 & 52905.1 & $+138.8$ & 3500-7000; 5200-9200\\
\hline
\multicolumn{4}{l}{\rlap{*}\ \ Relative to the epoch of $B$ maximum JD = 2452766
.3}
\end{tabular}
\label{tab3}
\end{table}

\section{The $UBVRI$ light curves}
\begin{figure*}
\centering
\includegraphics[width=14cm]{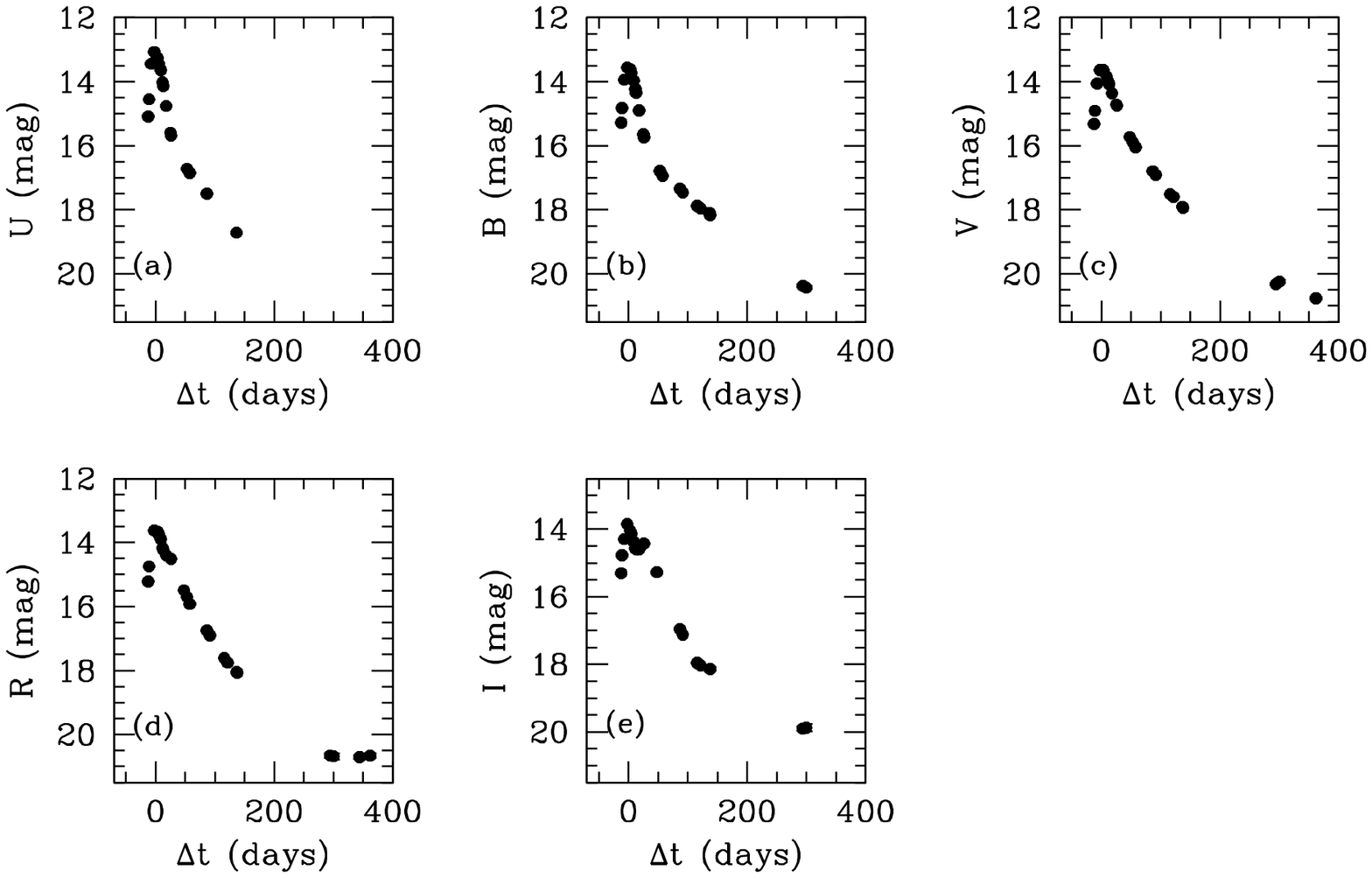}
\caption{The $UBVRI$ light curves of SN 2003du. The abscissae correspond to 
days since $B$ maximum. The errorbars in magnitude are less than or comparable 
to the point size.}
\label{fig2}
\end{figure*}

The $UBVRI$ light curves of SN 2003du are presented in Figs. \ref{fig2}a-e.  
Based on a cubic-spline fit to the points around maximum, it is estimated that 
the supernova reached a maximum $B$ magnitude of $13.53\pm 0.02$~mag on 
JD~$245\,2766.3\pm 0.5$. Following Phillips (\cite{p93}), we measure 
$\Delta m_{15}(B)=1.04\pm 0.03$. The main photometric parameters based on the 
light curves are listed in Table 4. 
The $V$ maximum occurred later than the $B$
maximum, while the maxima in $U$ and $I$ occurred $\sim 1$ day earlier, and
the maximum in $R$ coincided with the $B$ maximum. The $I$ light curve shows a
secondary maximum $\sim 25$~d after $B$ maximum at a magnitude $\sim 0.5$~mag
fainter than the first maximum, similar to `normal' SNe Ia (see e.g.\ Salvo
et al.\ \cite{s01}). A noticable rise in the $R$ band is also seen at similar 
times, although coverage in our data at that period is poor.

\begin{table*}
\caption{Photometric parameters of SN 2003du}
\begin{tabular}{lrrrrr}
\hline\hline
Data & \multicolumn{1}{c}{U} & \multicolumn{1}{c}{B} & \multicolumn{1}{c}{V} & 
\multicolumn{1}{c}{R} & \multicolumn{1}{c}{I} \\
\hline
epoch of max\rlap{*} & $765.0\pm 1.0$ & $766.3\pm 0.5$ & $767.3\pm 0.5$ &
$766.6\pm 0.5$ & $764.4\pm 0.5$\\
magnitude at max & $13.06\pm 0.03$ & $13.53\pm 0.02$ & $13.61\pm 0.02$ &
$13.61\pm 0.02$ & $13.85\pm 0.02$\\
early decline rate$^{\dagger}$ & 12.0 & 10.6 & 4.5 &\\
decline rate days 100-300$^{\dagger,+}$ & & 1.45 & 1.51 & 1.66 & 1.07\\
$\Delta m_{15}(B)$ & & $1.04\pm 0.03$ & \\
\\
colours at B max & & \multicolumn{1}{c}{$U-B$} & \multicolumn{1}{c}{$B-V$} & 
\multicolumn{1}{c}{$V-R$} & \multicolumn{1}{c}{$R-I$}\\
                 & & $-0.47\pm 0.04$ & $-0.085\pm 0.03$ & $-0.017\pm 0.03$ &
$-0.274\pm 0.03$\\
\hline
\multicolumn{6}{l}{\rlap{*}\ \ \ JD 2452000+}\\
\multicolumn{6}{l}{$^{\dagger}$\ \ mag (100d)$^{-1}$}\\
\multicolumn{6}{l}{$^+$\ \ Days since B maximum}\\
\end{tabular}
\label{tab4}
\end{table*}

\begin{figure*}
\centering
\includegraphics[width=14cm]{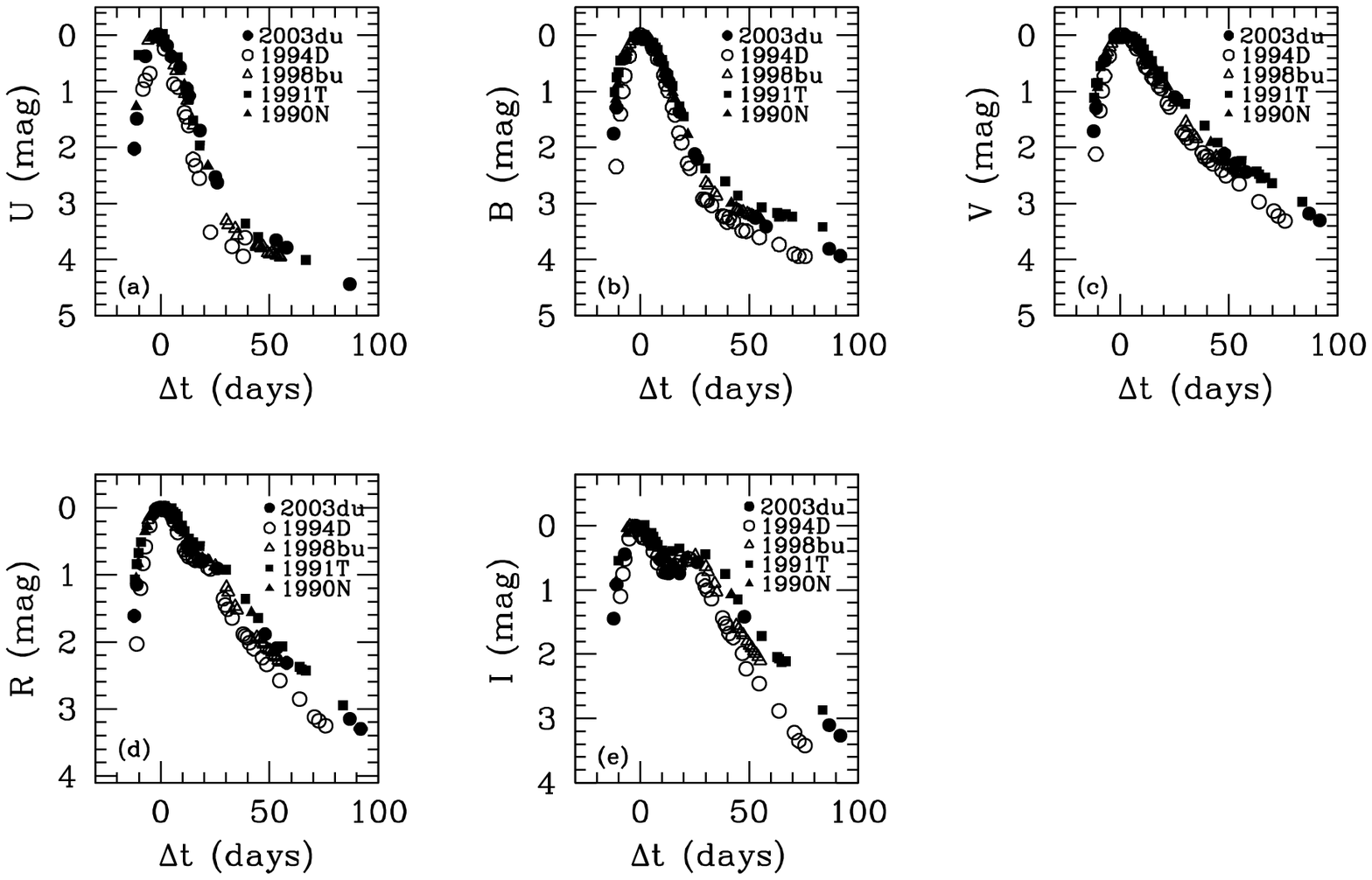}
\caption{$U$, $B$, $V$, $R$ and $I$ light curves of SNe 2003du, 1991T, 1990N, 
1998bu and 1994D. The ordinate in each panel is the magnitude below the 
respective maximum, and the abscissae represent the days since respective 
$B$ maximum} 
\label{fig3}
\end{figure*}

We discuss in the following the light curve in each bandpass. The light curves
in each bandpass are plotted individually, together with the light curves of
other well observed type Ia SNe: SN 1994D (Richmond et al.\ \cite{r95}), 
SN 1990N (Lira et al.\ \cite{li98}), SN 1998bu (Suntzeff et al. \cite{set99}) 
and SN 1991T (Lira et al. \cite{li98}). All light curves are plotted such that 
the individual peak brightness is scaled with respect to that of its respective
maximum (see Figs.\ \ref{fig3}a-e). 
The parameter $\Delta m_{15}(B)$ for each of these SNe are
$\Delta m_{15}(B) = 1.26$ (SN 1994D: Patat et al.\ \cite{pa96}), 
$\Delta m_{15}(B) = 1.03$ (SN 1990N: Lira et al.\ \cite{li98}),
$\Delta m_{15}(B) = 1.01$ (SN 1998bu: Suntzeff et al. \cite{set99}) and
$\Delta m_{15}(B) = 0.95$ (SN 1991T: Lira et al.\ \cite{li98}).

The $U$ band light curve (Fig. \ref{fig3}a) indicates that SN 2003du is broader
than SN 1994D, but narrower than SN 1991T. The light curve of SN 2003du is
however very similar to those of SN 1990N and SN 1998bu. A similar trend is
seen in the $B$ band also (Fig. \ref{fig3}b). SN 2003du rises to maximum slower
than SN 1994D, but faster than SN 1991T. The width of peak at 0.6 mag below
maximum for SN1991T is about 21 days, for SN1994D is about 16 days and for
SN2003du is about 20 days, similar to that of SN 1990N. Although differences
are seen in the early decline, the decline rate of 1.45 mag~(100)$^{-1}$
seen in SN 2003du at later phases is similar to that seen in 
SN 1991T and SN 1990N (Lira et al.\ \cite{li98}).

The $V$ band light curve of SN2003du follows the same trend as that in the
$U$ and $B$ bands. The early decline is slower than SN 1994D, but faster than
SN 1991T. The shoulder that is seen close to 25 days after $B$ maximum in the 
light curves of SN 1991T and SN 1998bu is not seen clearly in the light curve 
of SN2003du mainly due to the paucity of observation during that period.

A secondary peak characterizes the light curves of all normal and overluminous 
type SNe Ia, in both the $R$
and $I$ bands. The $R$ band light curve is similar to SN 1998bu until the 
secondary maximum. Subsequently, the decline appears slower than that of 
SN 1998bu. In the $I$ band, the light curve of SN 2003du is similar to that of
SN 1998bu until the onset of the secondary maximum. Although the period
around the secondary maximum is not well covered by our data, it appears that
the day of secondary maximum (since $B$ maximum) in SN 2003du is similar to 
that of SN 1991T. Also, the decline from day 50-100 since $B$ maximum follows 
that of SN 1991T rather than SN 1998bu. It is interesting to note here that 
the decline until the onset of the secondary maximum in SN 1990N closely 
matches that of SN 1991T.

From the light curves of SN 2003du in different filters and the estimate of
$\Delta m_{15}(B)$ it is evident that SN 2003du differs significantly from 
normal SNe Ia occurring in ellipticals (like SN 1994D), being more
similar to SN 1990N or SN 1998bu which occurred in spirals.

\section{The colour curves}
\begin{figure*}
\centering
\includegraphics[width=12cm]{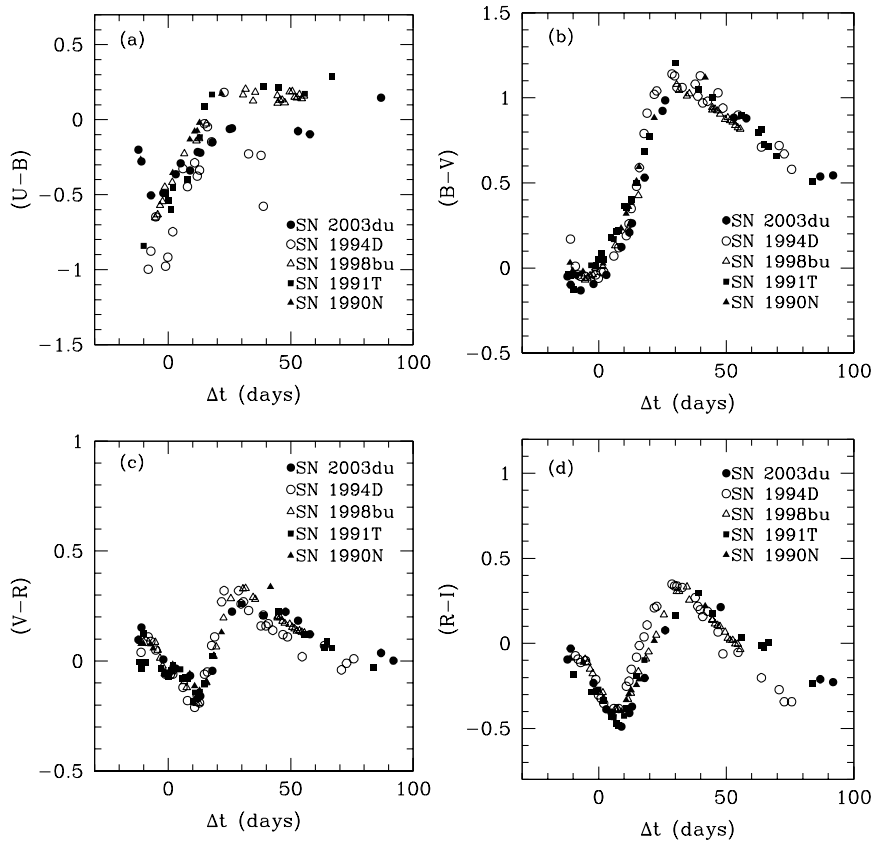}
\caption{The $U-B$, $B-V$, $V-R$ and $R-I$ colour curves of SN 2003du. Also 
shown for a comparison are the colour curves of SNe 1991T, 1990N, 1998bu and 
1994D. The abscissae correspond to days since respective $B$ maximum.}
\label{fig4}
\end{figure*}

The dereddened $(U-B)$, $(B-V)$, $(V-R)$, and $(R-I)$ colour curves for 
SN 2003du are presented in Figs. \ref{fig4}a-d, together with those of 
SN 1991T, SN 1990N, SN 1998bu and SN 1994D. SN 2003du is corrected only for
the Galactic reddening of $E(B-V)=0.01$ (Schelgel et al.\ \cite{s98}), and the 
host 
galaxy extinction is assumed to be zero (refer to next section). The colours of
the SNe Ia used here for comparison were reddening corrected using the Cardelli 
et al (\cite{cd89}) extinction law and the $E(B-V)$ values of $E(B-V) = 0.13$ 
for SN 1991T (Phillips et al.\ \cite{pet92}), $E(B-V) = 0.32$ for SN 1998bu 
(Hernandez et al.\ \cite{he00}), $E(B-V) = 0.04$ (Richmond et al. \cite{r95}). 
SN 1990N has been corrected using only the Galactic reddening value of 
$E(B-V) = 0.026$ (Schelgel et al.\ \cite{s98}).

The $(U-B)$ colour (Fig. 4a) of SN 2003du is significantly different from that 
of SN 1994D, while the overall trend matches those of SNe SN 1991T, SN 1990N 
and SN 1998bu. At 12 days before $B$ maximum, the $(U-B)$ colour of SN 2003du 
is about $-0.2$, and gets bluer, reaching a value of $-0.47\pm 0.03$~mag at $B$ 
maximum. This value compares well with that of the type Ia SN 1996X (Salvo et 
al.\ \cite{s01}). SN 2003du is redder compared to SN 1991T and SN 1998bu before 
and at $B$ maximum. Subsequently, it remains bluer than SN 1991T, SN 1998bu and 
SN 1990N.

The $(B-V)$ colour (Fig. 4b) of SN 2003du follows the general trend of the
colour curves of other type Ia supernovae. At 12 days before $B$ maximum, 
the $(B-V)$ colour
of SN 2003du is $-0.05$~mag, which gets bluer, reaching minimum $-0.13$~mag 
at 7 days before $B$ maximum. At $B$ maximum, $(B-V)$ is $-0.08$~mag, which
is consistent with the intrinsic colours at maximum observed in other SNe Ia,
which are in the range $\approx -0.1\,-\,+0.1$~mag. Similar to the other SNe Ia 
SN 2003du attains $(B-V)=1$~mag $\sim 30$ days after the $B$ maximum, and 
thereafter the colour gets bluer. During pre-maximum to 15 days after $B$ 
maximum phase, SN 2003du is bluer than the other supernovae in comparison, 
subsequently  the $(B-V)$ colour compares well with SN 1991T.

As in the case of the $(U-B)$ and $(B-V)$ colour curves, the $(V-R)$ (Fig. 4c)
colour 
curve of SN 2003du also evolves in a manner similar to other SNe Ia. While
some differences are seen in the pre-maximum $(V-R)$ evolution of SN 2003du and
SN 1991T, the evolution of the $(V-R)$ evolution of SN 2003du after the $B$
maximum is very similar to that of SN 1991T. The $(V-R)$ colour is bluest at
$-0.17$~mag 12 days past $B$ maximum, and reaches 0.22~mag between 25--45 days
after the $B$ maximum.

The $(R-I)$ (Fig. 4d) colour evolution of SN 2003du is very similar to that of 
SN 1991T and generally bluer than the other SNe Ia being compared here. The
$(R-I)$ colour reaches its bluest value of $-0.5$~mag around 10 days since
$B$ maximum, and subsequently gets increasingly redder. At later phases, (beyond
day 50), both SN 2003du and SN 1991T are redder than the other SNe Ia.

\section{The absolute magnitude of SN 2003du}

The absolute magnitude of SN 2003du can be derived from the distance to the host
galaxy or using the relation between the light curve shape and the SN absolute
magnitude.

The reddening
due to extinction within the host galaxy may be estimated either using the
$(B-V)$ colour at maximum light, or using the $(B-V)$ colour during the
decline, at a reference point of $t_V=53$, where $t_V$ is the
phase measured in days since $V$ maximum (Phillips et al.\ \cite{pet99}).
The observed $(B-V)$ at maximum (given $(B-V)_{\rm{max}}=0.07$; Phillips et
al.\ \cite{pet99}) suggests a reddening which is virtually zero,
while the observed $(B-V)$ colour at $t_V=53$ gives a reddening estimate of
$E(B-V)=0.07\pm 0.05$. Based on these estimates, we assume zero reddening due 
to the host galaxy, a result also confirmed by the absence of the Na I D 
absorption features in the spectra. 

Using the relation between $M_B$ versus $\Delta m_{15}(B)$ to calibrate the
SN absolute magnitude, we obtain $M_B^{\mathrm{max}}=-19.30\pm 0.2$ 
(Hamuy et al.\ \cite{h96a}) and $M_B^{\mathrm{max}}=-19.31\pm 0.2$ (Phillips 
et al.\ \cite{pet99}). Both methods give the same value for the absolute 
magnitude, which corresponds to a distance modulus of $\mu = 32.83\pm 0.2$. 
Della Valle et al. (\cite{d98}) give an improved estimate to the intercept in
the Hamuy et al.\ (\cite{h96a}) relation, using which we obtain
$M_B^{\mathrm{max}}=-19.42\pm 0.3$ and $\mu = 32.95\pm 0.3$. In the following, 
we use the average distance modulus of $\mu = 32.89\pm 0.4$

\section{The bolometric light curve}
The bolometric light curve of SN 2003du is estimated using the optical 
observations presented here, integrating the flux emitted in the $U$, $B$, $V$, $R$ and $I$ passbands. During the early phase, only the days on which data
were available in all passbands were used. However, for
the late phase (beyond day 90), we have integrated only over the available
passbands. Contardo et al.\ (\cite{c00}) have shown that the errors due to 
missing passbands in the optical do not exceed 10\%. The bolometric light curve 
for a distance modulus $\mu=32.89$, using only the optical data presented here 
indicates a peak at $\log L=43.06$. 

\begin{figure}
\resizebox{\hsize}{!}{\includegraphics{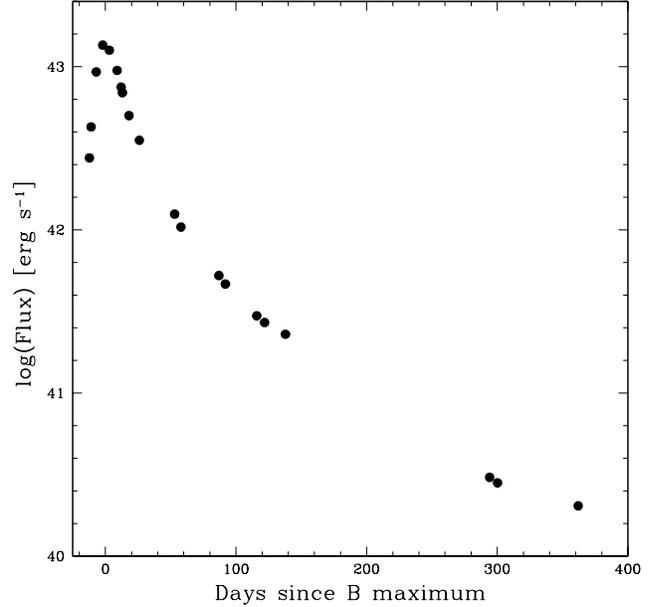}}
\caption{The bolometric light curve of SN 2003du.}
\label{fig5}
\end{figure}

Combining the ultraviolet and near-IR data with the optical data for SN 1992A, 
Suntzeff (\cite{s96}) has computed the {\it{uvoir}} bolometric luminosity, which
integrates the flux emitted in the range 2000~\AA - 2.2~$\mu$m. The ultraviolet
flux contributes about 10 percent (maximum) to the total {\it{uvoir}} 
luminosity, while the near-IR contribution is also a maximum of 10-15 percent
of the total {\it{uvoir}} luminosity for over $\sim 80$ days since $B$ maximum. 
Thus, although the maximum contribution
to the {\it{uvoir}} luminosity comes from the optical data, a correction is 
required to be applied the bolometric luminosity estimated using the optical 
data alone. Adopting a constant, total contribution of 20 percent from both 
ultraviolet and the near-IR regions, the {\it{uvoir}} luminosity estimated here
indicates a peak luminosity of $\log L=43.14$. The time evolution of the 
{\it{uvoir}} bolometric luminosity is presented in Fig. \ref{fig5}.

The amount of $^{56}$Ni mass ejected may be estimated using the peak 
luminosity (Arnett \cite{a82}). Assuming the rise time of SN 2003du to be 
$\sim 18$~days
and using the luminosity obtained by the optical data alone, the amount of
$^{56}$Ni ejected is estimated to be $0.73\,{\rm{M}}_\odot$. Using the corrected
bolometric luminosity, the $^{56}$Ni mass estimate is $0.88\,{\rm{M}}_\odot$.

\section{The spectral evolution}
\begin{figure}
\resizebox{\hsize}{!}{\includegraphics{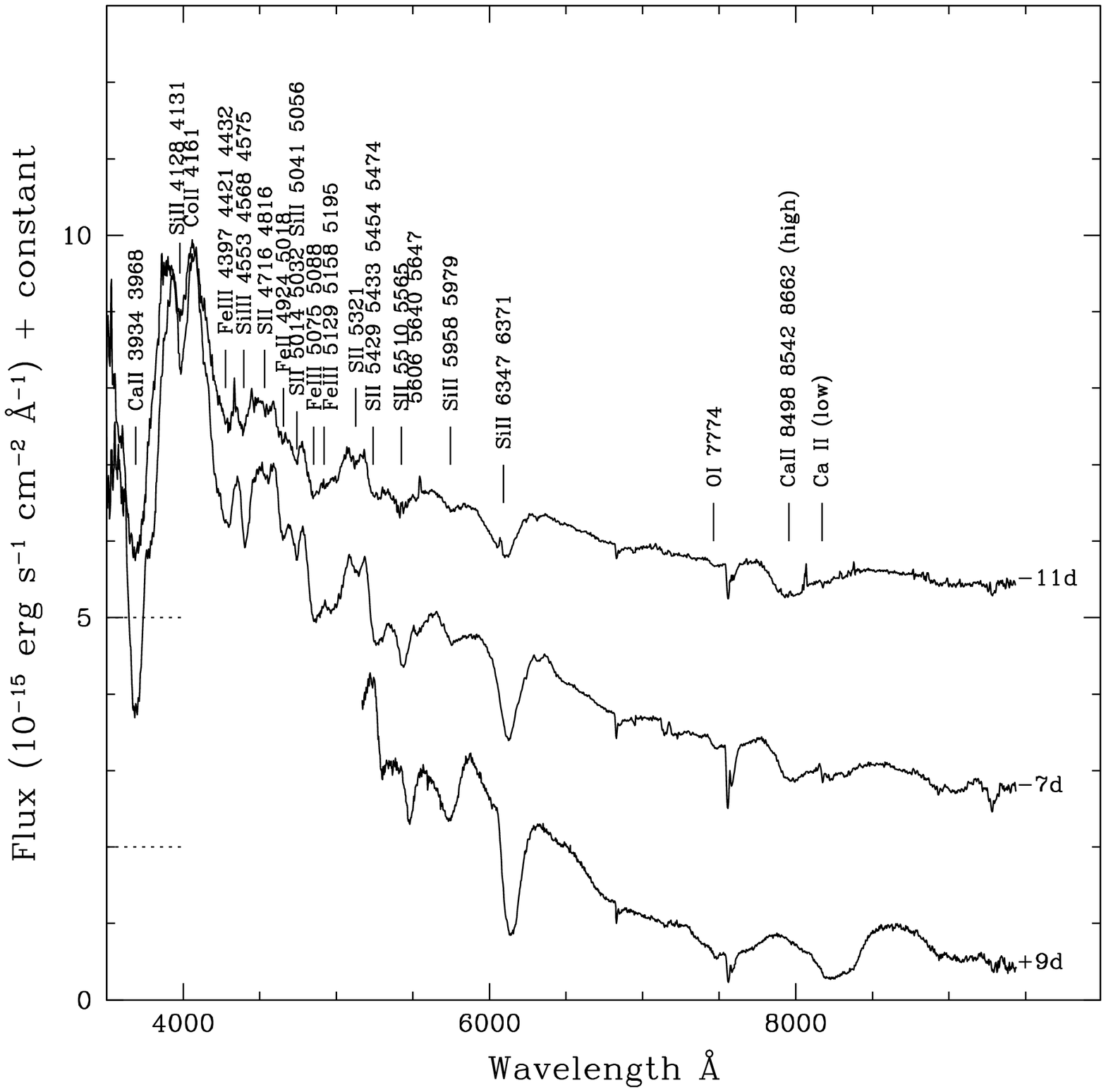}}
\caption{The spectral evolution of SN 2003du from phase $-11$d to $+9$d. 
The spectra are not corrected for reddening. For clarity, the spectra have been
displaced vertically. Dotted lines at the left indicate the zero flux level for
each spectrum. For +9d, zero flux is the $x$-axis. Wavelength scale is in the 
observer rest frame. Main line identifications are marked according to Mazzali 
et al.\ (\cite{mz93}).}
\label{fig6}
\end{figure}

\begin{figure}
\resizebox{\hsize}{!}{\includegraphics{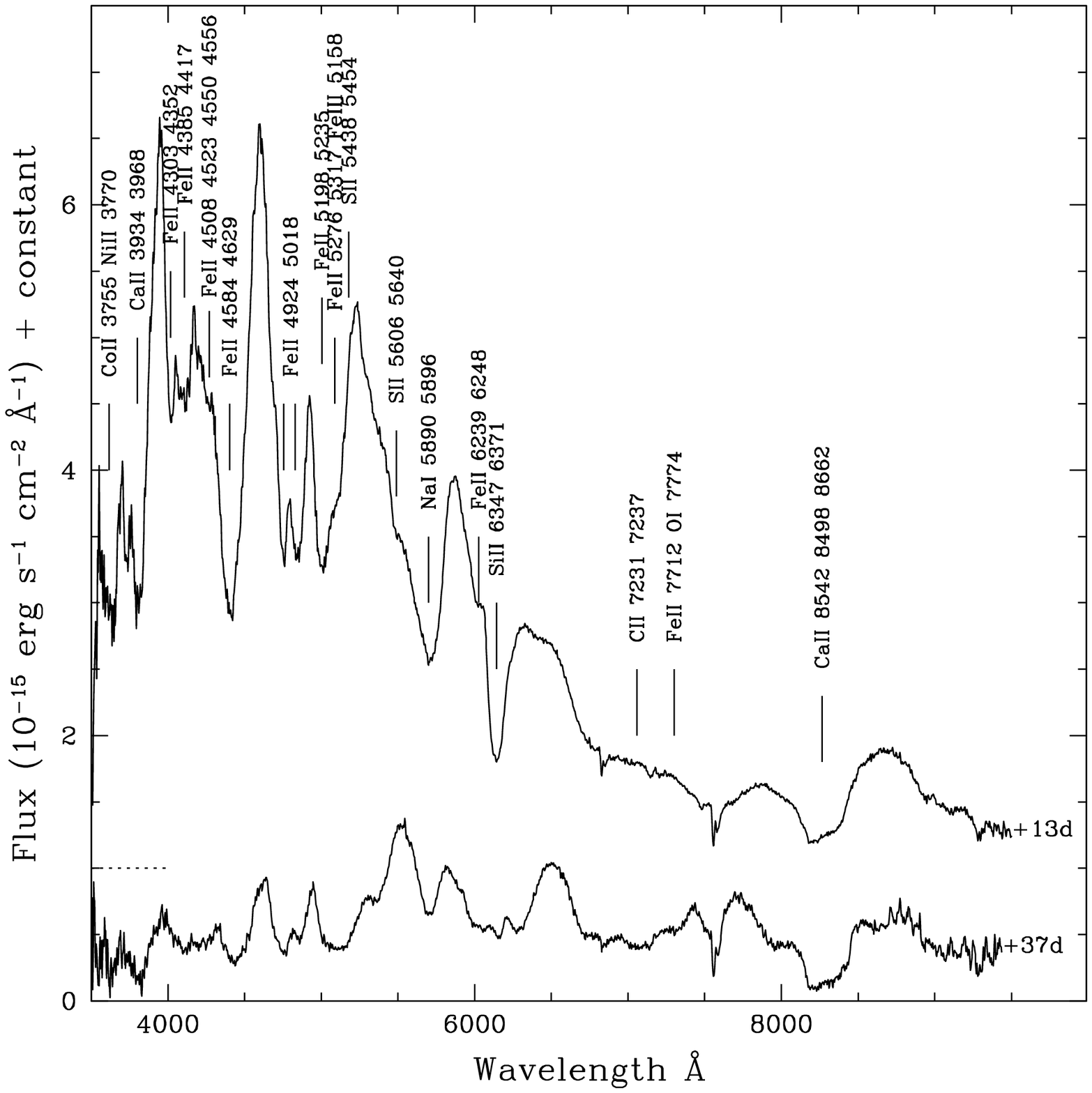}}
\caption{The spectral evolution of SN 2003du from phase $+13$d to $+37$d. 
The spectra are not corrected for reddening. For clarity, the spectrum of +13d 
has been displaced vertically and the dotted line at the left indicates the 
zero flux level. The zero flux for +37d is the $x$-axis. Wavelength scale is 
in the observer rest frame. Main line identifications are marked according to 
Mazzali et al.\ (\cite{mz93}).}
\label{fig7}
\end{figure}

\begin{figure}
\resizebox{\hsize}{!}{\includegraphics{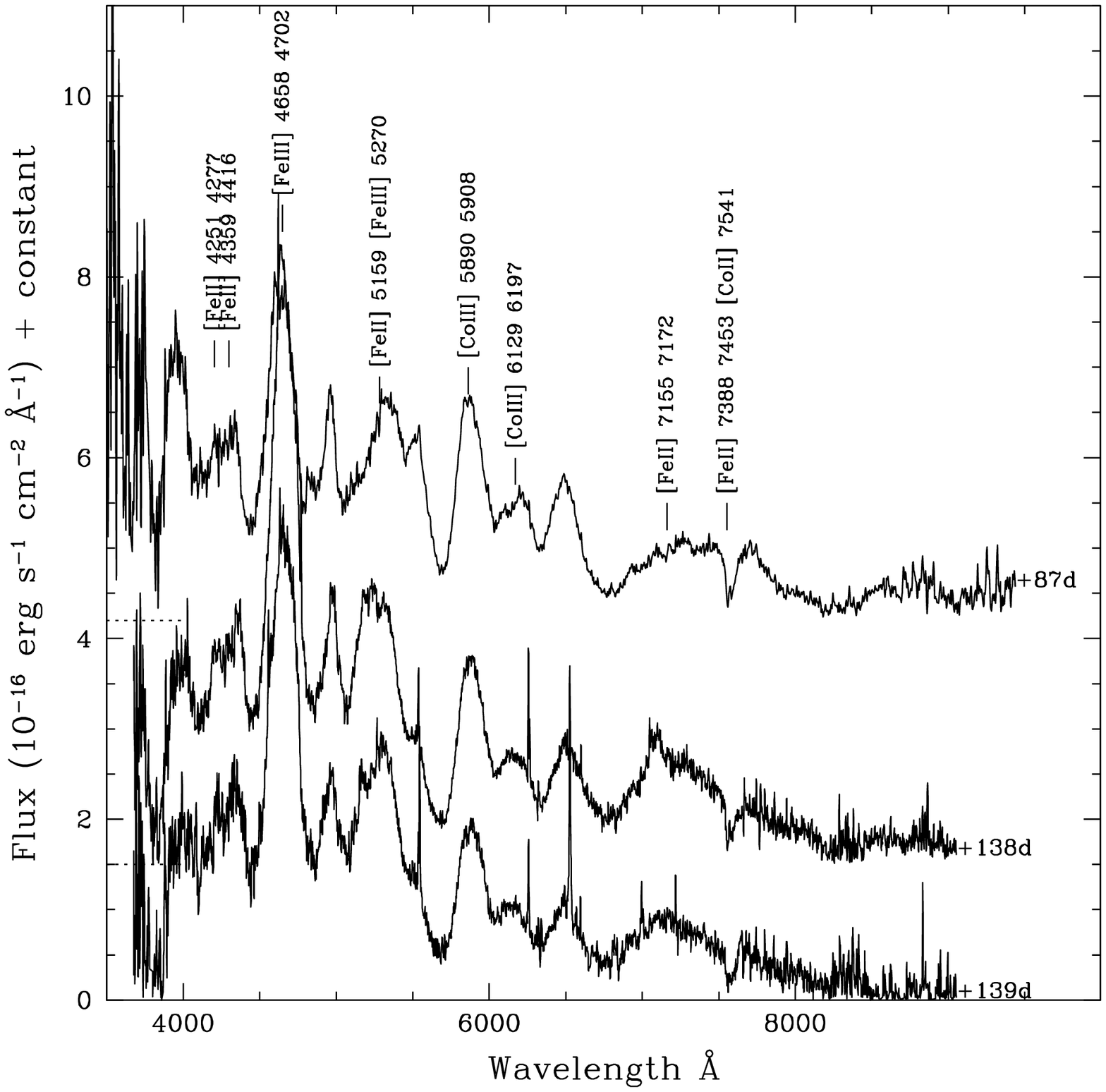}}
\caption{The spectral evolution of SN 2003du from phase $+87$d to $+139$d. 
The spectra are not corrected for reddening. For clarity, the spectra have been
displaced vertically. Dotted lines at the left indicate the zero flux level for
each spectrum. For +139d, zero flux is the $x$-axis. Wavelength scale is in 
the observer rest frame. Main line identifications are marked according to 
Bowers et al.\ (\cite{bet97}).}
\label{fig8}
\end{figure}

Spectra of SN 2003du have been obtained at 4 epochs, from phase $-11$ to $+139$.
They are presented in Figs. \ref{fig6}, \ref{fig7} and \ref{fig8}. The early 
spectra are
characterized by the deep absorption trough around 6000~\AA, due to Si\,II,
typical of ``normal'' SNe Ia.

The spectral evolution of SN 2003du is similar to that of other SNe Ia, such as
SN 1998bu (Jha et al.\ \cite{j99}), SN 1990N (Leibundgut et al.\ \cite{le91}), 
SN 1996X (Salvo et al.\ \cite{s01}) and SN 1994D (Patat et al.\ \cite{pa96}). 
At early phases, the
continuum is blue and dominated by lines due to the Fe-group and intermediate
mass elements -- Si, Ca, S, Mg. All lines exhibit the characteristic P-Cygni
profiles, with the absorption shifting towards longer wavelengths, indicating
a decrease in the expansion velocities with time. However, it is also possible
that we are observing different layers that are moving at different velocities.
The spectra at the later 
phases indicate a redder continuum, with nebular lines due to [Fe\,II] and 
[Fe\,III], which get dominant with time. 

\begin{figure}
\resizebox{\hsize}{!}{\includegraphics{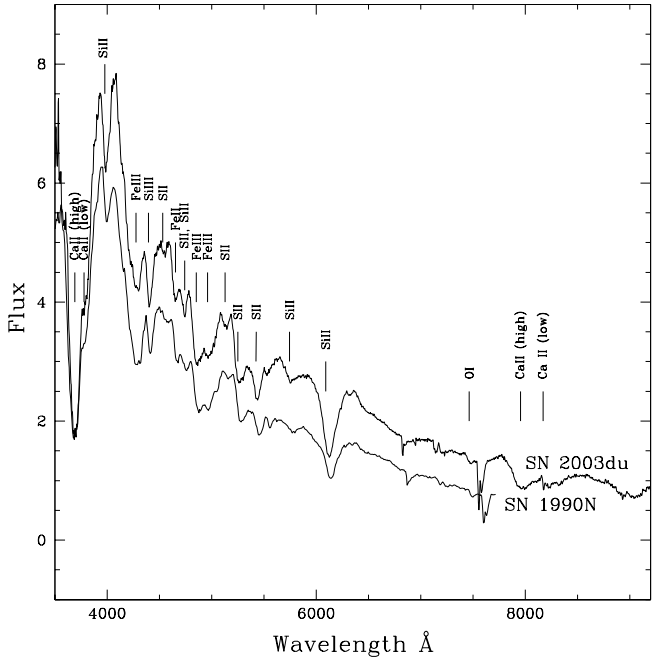}}
\caption{Comparison between the observed spectra of SN 2003du and SN 1990N
on day $-7$. The spectrum of SN 1990N was obtained from 
{\it{http://hej3.as.utexas.edu/$\sim$www/SN/spec/sn1a/90N/}}. The flux units are
arbitrary, and the wavelength scale is corrected for the recession velocity of 
the respective parent galaxy.}
\label{fig9}
\end{figure}

A more detailed comparison shows that although the spectra of all SNe Ia are
similar, there are some differences in the detailed shapes and velocities of
the features. For instance, in the early phase, the Ca II (H \& K) feature in 
SN 2003du is stronger than in other SNe Ia being compared here. A similar 
strength of the Ca II (H \& K) feature is seen in the $-7$~d spectrum of
SN 1990N also (Leibundgut et al.\ \cite{le91}). In Fig. \ref{fig9}, we plot the
$-7$~d spectrum of both SN 2003du and SN 1990N. The similarity between the
two SNe is remarkable. The strength of the O\,I 7774~\AA\ feature in SN 2003du
is also more compared to the other SNe. 

The high velocity absorption
component of the Ca II infrared triplet detected by Gerardy et al.\ (\cite{g04})
in the spectra of SN 2003du between day $-5$ to day $+2$ is seen in the spectra 
of days $-11$ and $-7$ presented here. This high velocity component is detected
in the Ca II (H \& K) absorption also. The low velocity component of the Ca II
features appeared weakly on day $-7$. While it is clearly detected in the IR
triplet feature, it appears as a slight inflexion in the (H \& K) feature. A
similar inflexion is seen in the day $-7$ spectrum of SN 1990N also (see Fig. 
\ref{fig9}). The high velocity component disappeared by day $+6$. Gerardy 
et al.\ (\cite{g04}) attribute
this high velocity component as being caused by a dense shell formed when 
circumstellar material is overrun by the rapidly expanding outermost layers
of the SN ejecta. 

\begin{figure}
\resizebox{\hsize}{!}{\includegraphics{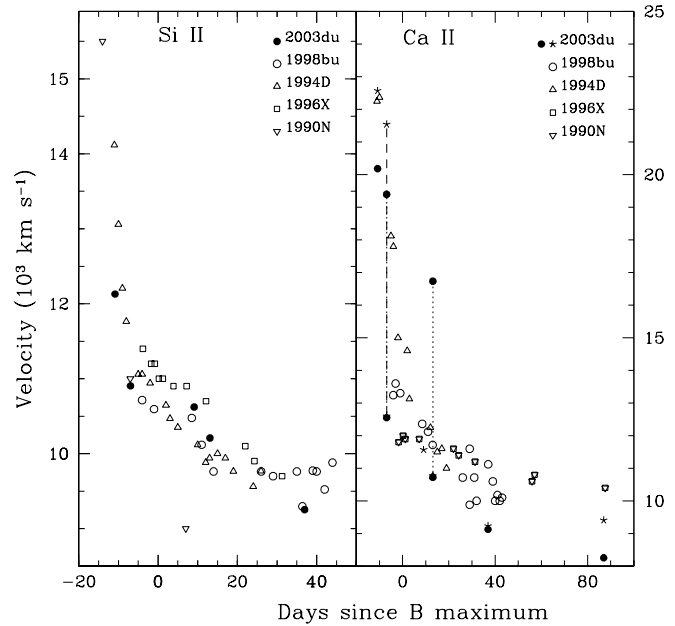}}
\caption{Expansion velocities derived from the absorption minima of Si II 
6355~\AA\ and the Ca II H\,\&\,K for SN 2003du, SN 1994D, SN 1990N 
and SN 1996X (see text for references). The Ca II IR-triplet absorption 
velocities are also plotted for SN 2003du ({\it{stars}}). The expansion 
velocities have been corrected for the 
recession velocity of the host galaxy. The vertical lines connect the low and
the high velocity components of the Ca II absorption in SN 2003du when both
components were detected.}
\label{fig10}
\end{figure}

A quantitative comparison of
the Si II 6355~\AA\ and the Ca\,II (H\,\&\,K) feature absorption velocities 
between SN 2003du and other typical SNe Ia: SN 1994D (Patat et al.\ 
\cite{pa96}), SN 1998bu (Jha et al.\ \cite{j99}), SN 1996X (Salvo et al.\ 
\cite{s01}) and SN 1990N (Leibundgut et al.\ \cite{le91}) is illustrated in 
Fig.\ \ref{fig10}. From the Figure, it is seen that the velocities of the 
Si\,II feature in the pre-maximum phase
in SN 2003du is lower than that of SN 1994D, but very similar to that of 
SN 1998bu and SN 1990N. A similar trend is seen in the high velocity component
of the Ca II (H\,\&\,K) absorption. The evolution of the absorption velocities 
in the post-maximum phase is very similar to the other SNe Ia and falls well
within the scatter defined by the other objects. 

Spectroscopically, SN 2003du appears to be a normal type Ia in a spiral galaxy.

\section{Summary}

The $UBVRI$ photometry and optical spectra of SN 2003du obtained over a period
of nearly a year are presented here. Based on the observed light curve, we
estimate SN 2003du reached a maximum in the $B$ on JD$\,245\,2766.3\pm 0.5$ 
with an apparent magnitude of $B=13.53\pm 0.02$~mag, and colour 
$(B-V)=-0.08\pm 0.03$~mag. The luminosity decline rate,
$\Delta m_{15}(B)=1.04\pm 0.04$~mag indicates an absolute $B$ magnitude at 
maximum of $M_B=-19.34\pm 0.3$ and the distance modulus to the parent galaxy as
$\mu=32.89\pm 0.4$.
SN 2003du is definitely brighter than SN 1994D (Patat et al.\
\cite{pa96}), which is prototypical of SNe Ia occurring in early type galaxies.
The light curve shapes are however, similar, though not identical, to those of 
SNe 1998bu and 1990N, both of which had luminosity decline rates similar to 
that of SN 2003du and occurred in spiral galaxies.
The peak bolometric luminosity indicates that $\sim 0.9\,{\rm{M}}_\odot$ mass
of $^{56}$Ni was ejected by the supernova.

The spectral evolution and the evolution of the Si II and Ca II absorption 
velocities closely follows that of SN 1998bu and SN 1990N, and in general, is 
within the scatter of the velocities observed in normal type Ia supernovae. 
The spectroscopic and photometric behaviour of SN 2003du is quite typical for
SNe Ia in spirals. 

The spectra presented here cover phases earlier than those covered by
Gerardy et al.\ (\cite{g04}), and the high velocity absorption component 
in the Ca II (H \& K) and IR-triplet features is detected as early as $-11$ 
days before $B$ maximum. While the first spectrum obtained by us on day $-11$
shows only the high velocity absorption component, the low velocity component
appears on day $-7$.

\begin{acknowledgements}
We thank all the observers of the 2-m HCT who kindly provided part of their
observing time for the supernova observations. We thank the referee M. Della 
Valle for his prompt and useful comments and suggestions. This work has made 
use of
the NASA Astrophysics Data System and the NASA/IPAC Extragalactic Database 
(NED) which is operated by the Jet Propulsion Laboratory, California Institute 
of Technology, under contract with the National Aeronautics and Space 
Administration. We also thank the Supernova group at the Department
of Astronomy, University of Texas, Austin, USA for posting the data on SN 1990N
on the {\it{web}}.
\end{acknowledgements}

\end{document}